# On the Control of a Leading-Edge Vortex & its Liftoff on a Cranked, Swept Back Wing

**H. Kalyankar[(1)], L. Taubert[(1)] & I. Wygnanski[(1)]**

[(1)] Aerospace and Mechanical Engineering department, University of Arizona, 1130 N. Mountain Ave, Tucson, AZ-85721, USA, Email: wygy@arizona.edu

## ABSTRACT

This study examines flow over a cranked λ-wing model with a sweep of $\Lambda=60°$ of the inboard leading edge (LE) that changed to $\Lambda=30°$ outboard of the crank. The study focuses on the liftoff of the inboard Leading-Edge Vortex (LEV) and its influence over the flow on the outer wing. Stereoscopic Particle Image Velocimetry (SPIV) indicates that the flow is mostly attached in the crank's vicinity at incidence $α=12°$ while being largely separated and dominated by strong outboard spanwise flow at $α=14°$. The separation is mitigated by a small, steady supersonic jet that interacted with the lifted vortex and changed the pitch characteristic of the entire model. The form of this interaction is still unknown (i.e. does it alter the character of the LEV or mostly affect its path) as this is a study in progress. The use of Proper Orthogonal Decomposition (POD) revealed the unsteadiness of the separated structures that may cause wing flutter.

## 1. INTRODUCTION

The pursuit of air superiority translates to high-speed cruise with supersonic capability, agility, and stealth. It led to the design of slender, thin, swept-back wings with fuselages that blend with the wings, and to unmanned combat air vehicles that are mostly tailless, having either a λ or a cranked Δ shaped planforms. Since the leading-edge radii of most of these wings are less than 1% of the chord, they create Leading Edge Vortices (LEVs) at relatively low angles of incidence, and they rely in part on vortex lift when flying slower or manoeuvring [1]. When these vortices are seen by the naked eye due to natural condensation occurring in their low-pressure core, they appear to be steady even after lifting off from the wings of the aircraft that generates them.

Vortex flows on tailless configurations are complex and they involve interactions among multiple vortices even at low Mach numbers. Pressure distributions over these configurations appear to confirm the stationary behavior of the LEVs even when there is a plurality of them, and they interact with one another [2]. However, the reported pressure distributions were mostly measured using Pressure Sensitive Paint (PSP) and data acquisition procedures that are not sensitive to short temporal changes. This led to numerical simulations that focused on spatial resolution rather than temporal one. Many of the programs used three-dimensional solvers based on discretized Reynolds-averaged Navier-Stokes (RANS) equations. The validation of the results depended on the turbulence models used and on the mesh sizes employed. At times hybrid approaches were used that combined RANS with Large Eddy Simulations (LES) but the results were similar: a LEV detached itself from the leading edge (LE) due to a step-change in LE radius (SACCON model), moved downstream before turning in the spanwise direction near the trailing edge of the outer portion of the λ planform [3] [4]. Since the task was to predict integral forces and moments, the steadiness of the computations was not seriously questioned, although the nose-up pitch instability at high incidence angles was of concern.

The unsteadiness of a detached vortex is of great concern when it either impinges on or attaches itself to a vertical stabilizer creating buffet [5]. The pressure oscillations were attributed to the helical motion of a detached vortex following its breakdown. Although vortex breakdown is caused by its core's axial velocity component being brought to stagnation by the adverse pressure gradient associated with high incidence angle, it usually results in the detachment of the vortex from the surface to be followed by a meandering path downstream. LE strakes and canards generate strong vortices that improve the effectiveness of a rudder by inducing a downwash on it at high incidence angles, but the strong pressure oscillations result in flutter and fatigue. Accepting that the detached vortex contains high amplitude, coherent, quasi periodic, structures, leads one to ask if they originated upon detachment or they were embedded in the interior of the attached and stationary LEV. A LEV is bound by a curved shear layer and a solid surface. The shear layer is highly unstable, and it generates large Kelvin-Helmholtz (K-H) eddies that might be absolutely unstable when viewed normally to the LE, but it sheds some of its vorticity outboard parallel to the span, due to the sweep-back. The magnitude of the transport along the span is determined by the sweep angle, $Λ$, the distance from the root of the wing and the angle of incidence, $α$. Thus, on a given delta wing, an equilibrium condition may be attained for a single $α$. At smaller $α$ there is no LEV, while at larger $α$ it expands with increasing distance from the apex and eventually breaks down and detaches itself from the surface. The presence of a crank generates an adverse pressure gradient along the span that breaks this equilibrium and results in a premature liftoff of the LEV. Such a liftoff might occur without a crank but at a larger distance from the root of the wing or a larger $α$. The location of the LEV liftoff can also be triggered by a fence, a snag or even a discontinuity in the LE radius of the wing. A correlation between a LEV and K-H instability was established in [6] but only its effect on the integral forces and the size of the LEV were published in [7]. Since Margalit [6] provides a guide to the present research some of the unpublished results were included as background.

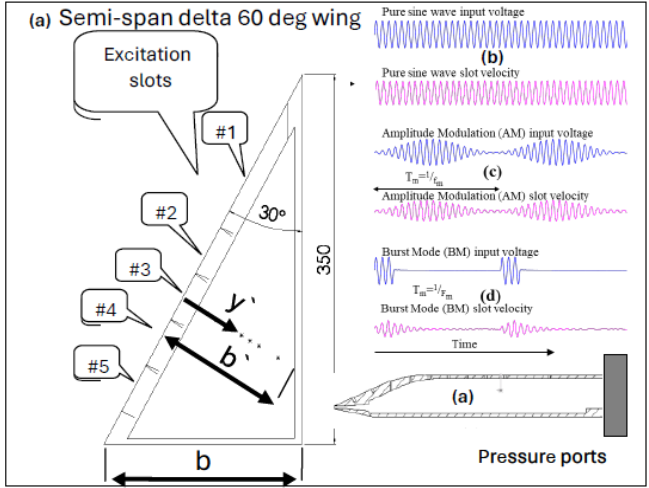 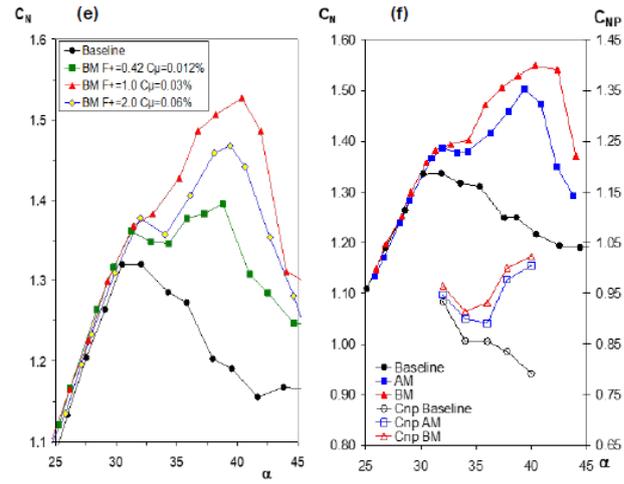

Figure 1:(a) The semi span delta wing model; (b-d) Leading edge slot calibration for different modes of input; (e) The variation of $C_N$ with $α$ for different burst mode frequencies (duty cycle =7.5% of time); (f) Comparing the effect of burst mode at $F^+$=1.0 & $C_μ$=0.03% with amplitude modulation of $F^+$=2.0 & $C_μ$=0.41%, including the effect on integrated pressure.

Experiments were performed on a semi-span delta wing model having a sweep-back, $Λ$= 60° and a sharp, beveled LE as shown in figure 1-a. The LE contained a 1mm wide segmented slot along its span. Five internally mounted piezo-electric actuators provided zero mass-flux periodic excitation to each of the 5 slots. The first slot near the apex was 34% of the LE in length while the other 4 were 12% each. Since the resonance frequency of the actuators was much too high the input voltage was either amplitude modulated (figure 1-c) or supplied in intermittent bursts of the carrier frequency (figure 1-d). Each actuator was calibrated by a hot wire and the calibration was expressed by oscillatory momentum coefficient, $C_μ$, as defined in [7]. The model was mounted on a balance and the flow was interrogated by a two-dimensional Particle Image Velocimeter (PIV) in a plane normal to the upper surface of the wing and the wind tunnel sidewall. There was also an array of pressure ports located on the upper surface downstream of the 3rd actuator slot. The tests were carried out at Reynolds number, $Re$ <0.36x10$^6$ based on the root chord of the model. More details about the experiment can be found in [7].

The dependence of the normal force, $C_N$, on the angle of attack is plotted in Figs. 1e for various normalized frequencies, $F^+$, provided by the burst mode. The highest value of $C_N$ was obtained for $F^+$=1 that increased $C_N$, from its maximum baseline value of 1.3 to $C_N$ =1.5 while increasing the stall angle from 30° to 42°. Doubling the frequency while doubling the $C_μ$ value attained an inferior result, suggesting that $F^+$ might have a larger influence on the flow than $C_μ$. The burst mode of actuation seems to be more effective than the amplitude modulation mode (figure 1-f) despite the large difference in $C_μ$, that is sensitive to the duty cycle. It is possible that a short pulse creates a wave packet that may best interact with local instabilities. According to [7], increasing $C_μ$ while using amplitude modulation at $F^+$=2 does not monotonically increase $C_N$. In fact, for 0.4%< $C_μ$<1.5%, an increase in $C_μ$ is deleterious, suggesting perhaps that the imposed oscillation couples with different mode of instability.

Figure 2 represents a slightly inclined component of the averaged streamwise vorticity, out-of-plane that is perpendicular to the wind-tunnel wall and to the wing's upper surface. The baseline data acquired at $x/c$=0.6 do not possess the characteristic features of a typical LE vortex; rather, the flow resembles a detached shear layer with a massive region of almost stagnant flow between the separated shear layer and the wing surface. With excitation, the shear layer is deflected toward the wing surface, reestablishing a vortical structure. This is accompanied by a near-wall counterclockwise (positive) vorticity (elaborated in figure 3), typical of an attached boundary layer flowing in the negative $y/b$ direction, that is, outboard. The differences between amplitude or burst modulation are small although it appears that the in-plane velocity contours are higher for the amplitude modulation mode.

The phase-averaged velocity and vorticity contours furnish some information about the periodic dynamics taking place inside the core of the LEV (figure 3). The bent shear layer near the LE of the wing (i.e. y/b<0.1 and z/b<0.2) retains approximately a constant vorticity throughout the cycle, however, as one follows this vortical layer toward the mid-span of the wing it becomes more diffused (i.e. broader) and less homogenous. At θ=0°, the blue contour terminates at y/b≈0.7 & z/b ≈0.4 and it reappears at full strength near the wing's surface (at y/b≈0.4 & z/b≈0.1). On the low-speed side of this location (marked by the brown circle) there is a core of an eddy. This is presumably where the mixing layer rolled into a K-H eddy that had undergone its maximum spatial amplification [8] [9]. At the same time there is another eddy (marked by a green circle) that represents the core circulation of the LEV. The size and intensity (clockwise rotation) of these two seems to be very different as shown by the in-plane streamlines on the lower left of figure 3. A quarter cycle later (θ=90°) most of the circulation in the LEV was transferred to upward and inboard to the vortex created by the shear layer and the vorticity within the shear layer surrounding the LEV is almost uniform along its entire periphery (at y/b ≈0.7, z/b ≈0.3). At θ=180° there is effectively only a single large vortex that moved closer to the surface during the time interval, but it remained at approximately the same spanwise location as at θ=90°. in the interval between 180°<θ<270° most of the circulation becomes diffused mostly due to convection in the downstream direction but also some dissipation must have reduced its intensity due to friction with the surface.

The in-plane, phase locked streamlines reveal that there are two vortex cores always present in the interior of the LEV (figure 3 bottom). A tight core (red) seen at θ=0° that is attached to the separated shear layer before it starts to bend toward the wing's surface, and a large core (green) that is associated with the recirculating flow near the surface. Since both vortices are rotating in a clockwise direction, they induce a clockwise rotation on one another, and the line connecting their cores rotates throughout the cycle. The vorticity in the tight core must have been strong enough to entrain the fluid from the large core in the short interval between 0°<θ<90° and become the dominant large vortex in the interior of the LEV. Thereafter, from 90°<θ<270° the large vortex moves toward the solid surface while losing intensity in the process, so at θ=270° both vortices are weak and of comparable strength. It is not known how much of this vorticity was transported out of the plane of interrogation in the direction of streaming and how much of it dissipated because only two-dimensional PIV was employed in the thesis [6].

Greenblatt et al [10] pursued the issue of vortex breakdown on a 60° delta wing using Dielectric Barrier Discharge (DBD) plasma actuators at much lower Reynolds numbers for potential use on Micro Air Vehicles. The increase in normal force that was obtained in [10] was very similar to that of [6] at a reduced frequency of $F^+$=1. Unfortunately, they too used a two-dimensional PIV system that simply confirmed the prior results, but they exposed the utility of the plasma actuation that was considered hitherto to be ineffective. The focus in [11] turned to forced reattachment of the flow to a flat plate airfoil using burst mode plasma actuation. The most effective reduced frequency that enclosed the smallest bubble upon the completion of flow reattachment was at $F^+<0.5$. The streamlines in the interior of such a bubble are shown in figure 4 and they clearly suggest that within the bubble there are two vortices that might have exchanged vorticity between themselves in a manner similar to the observations in figure 3. The extraction of vorticity from the mean flow to forced coherent structures and vice versa, as well as the exchange of vorticity between two vortex sizes created by fundamental and subharmonic frequencies were investigated in [12]. Coherent vorticity balance suggested that eddies whose tilt opposes the mean flow's vorticity extract vorticity from the mean flow and get amplified while eddies whose inclination is in the same direction as the mean velocity gradient, get stretched and decay. For this reason, small eddies amplify when the mixing layer is thin, while pairs of eddies do so further downstream (figure 4 top). The flow in a bubble may therefore be similar to the flow in a LEV, where two eddies continuously interact with the shear layer above them.

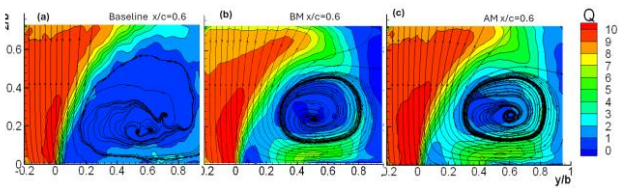

Figure 2: Total in-plane velocity contours and streamlines measured at x/c=0.6 in a plane perpendicular to the wing's surface at α=37.8°, Re=0.234 x 10^6. Excitation: AM: $F^+$=2.0, $C_\mu$=0.4%; BM: Duty cycle= 15%, $F^+$=1.0, $C_\mu$ =0.03%

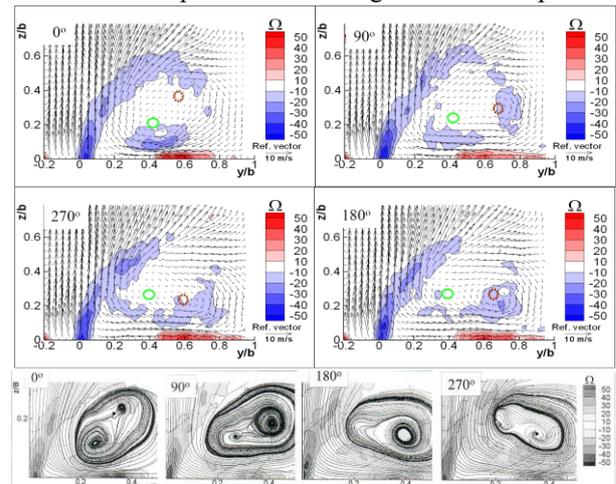

Figure 3: Top: Four phased locked velocity vectors and vorticity contours taken at conditions described in figure 2 for the amplitude modulated actuation. Bottom: In Plane streamlines for each of the 4 phases.

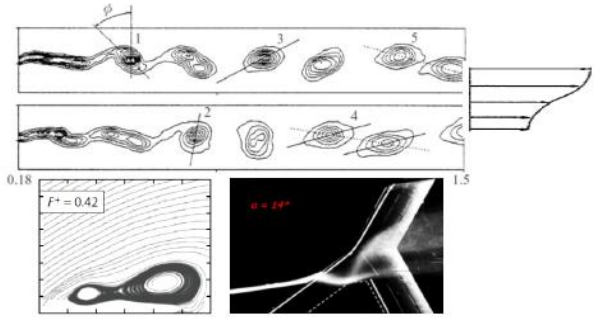
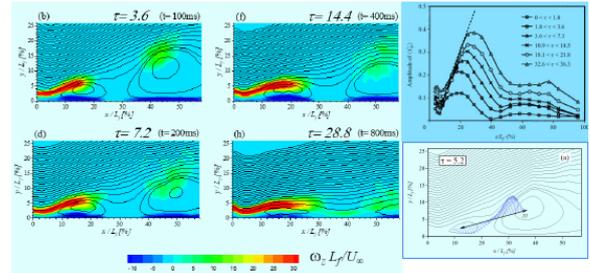

Figure 4: Top: Vorticity contours of the forced mixing layer [11], ____ fundamental; ………… subharmonic; Bottom: Effect of dimensionless frequency on time-averaged streamlines from PIV data on flat plate airfoil (Figure adapted with permission from Greenblatt et al. [11]) & smoke visualization on a coplanar joined wing model as seen by a naked eye.

Similar situations are encountered during the process of reattachment of the flow to a flat surface by periodic excitation [13]. Most effective reattachment takes place when 2 eddies are present over a surface of length $L_f$, thus for a given forcing frequency the amplified eddy covered half of the plate's length by the time it attained its maximum amplitude. When it continues from mid chord to the trailing edge it becomes larger, and it decays. The mean flow remains attached due to this equilibrium in vorticity transfer between the mean flow and the coherent K-H type eddies (figure 5). During the reattachment process fluid is transported from the surface to the free stream, thus the periodic removal of fluid from the surface lowers the pressure over the surface that forces the detached shear layer to change its direction toward the surface until it reattaches to it. The amplitude of the pressure oscillations on the surface increases as reattachment is approached (figure 5 top right).

Figure 5: Forced reattachment by periodic excitation [13]

When a LEV is lifted off the surface due to a crank, its dominant interior coherent eddies are lifted with it, and they may separate depending on their location within the LEV. This is what might have been observed in the smoke photograph obtained over a cranked, coplanar, joined wing shown in figure 4. The eddy close to the detached shear layer is still lifted and moves in the streamwise direction, while the eddy close to the surface may proceed outboard along the wing's span. Since a new LEV is usually created outboard of the crank, its interior structure may be affected by the periodicity of the lifted vortex, and this motivated the present research that may explain the strong buffet existing on the outer wing.

Among the available models to check the various hypothesis suggested above the semi-span SWIFT (Swept Wing Flow Test) was selected because it was extensively tested in numerous wind tunnels [14], including our own for the purpose of controlling pitch instability and buffet (figure 6-b). It has a large LE crank, located close to its mid-span with a short transition from the diamond shaped central body to a non-tapered outer wing. The airfoil sections are cambered, and an outwash is employed on the outer wing to delay and reduce the pitch instability. All the experiments were performed at the freestream velocity ($U_\infty$) of 25 m/s, resulting a Reynolds number ($Re$) of $1.2 \times 10^6$ based on the root chord.

## 2. RESULTS AND DISCUSSION

Veismann et. al. [14] investigated the effect of introducing a crank that changes the sweep back of the leading edge of the wing from 60° to 30° thus resulting in the creation of the LEV on the inboard part of the wing that is hardly noticeable by oil flow prior to liftoff due to the presence of the trip-strip. Initiation of LEV being inboard of the moment reference line (MRL) (broken red line in figure 6-b), generates a nose-up pitch departure for $C_L > 0.2$ (black curve with 'o' markers). Owing to the smaller sweep angle of the outer wing ($\Lambda=30°$), there is an increase in the chordwise velocity component, increasing the stagnation pressure, thus imposing a spanwise adverse pressure gradient for the inboard flow along the front attachment line. Hence, as the flow in the LEV interior approaches the crank, it lifts-off at $\alpha=13$-$14°$. Surface oil flow visualization at $\alpha=12°$ (figure 6-b inset, $C_L$~0.6) indicates that the surface streamlines near the Trailing Edge (TE) of the outer wing point outboard parallel to the TE. At $\alpha=14°$ (figure 6-b) a very large, lifted vortex footprint around the crank region is revealed. The focal point of the separation [15] appears to be just upstream of the crank location (marked with a red oval). The accumulation of oil might have been caused by trip-strip. The lift-off entrains fluid from the main element and directs it towards the outer wing. The evolution of the entrained flow on the outer wing as $\alpha$ is increased from $12°$ to $14°$ is noteworthy. At $\alpha=12°$, the spanwise outboard flow is attached. As the incidence angle increases slightly to $\alpha=13°$, the footprint of vortex lift-off starts showing up, with attached spanwise outboard flow, contributing to a nose-down pitch moment (picture available in [14]). At $\alpha=14°$, the lifted vortex is so strong that it entrains fluid on the surface of the outer wing and pulls it in the upstream direction, forming a saddle [15] that is hard to discern in in figure 6-b. This flow reversal induces partial flow separation on the control surfaces, resulting in a nose-up pitch moment. Concurrently, streamwise flow is maintained closer to the wingtip, thus creating a sinusoidal across the outer wing.

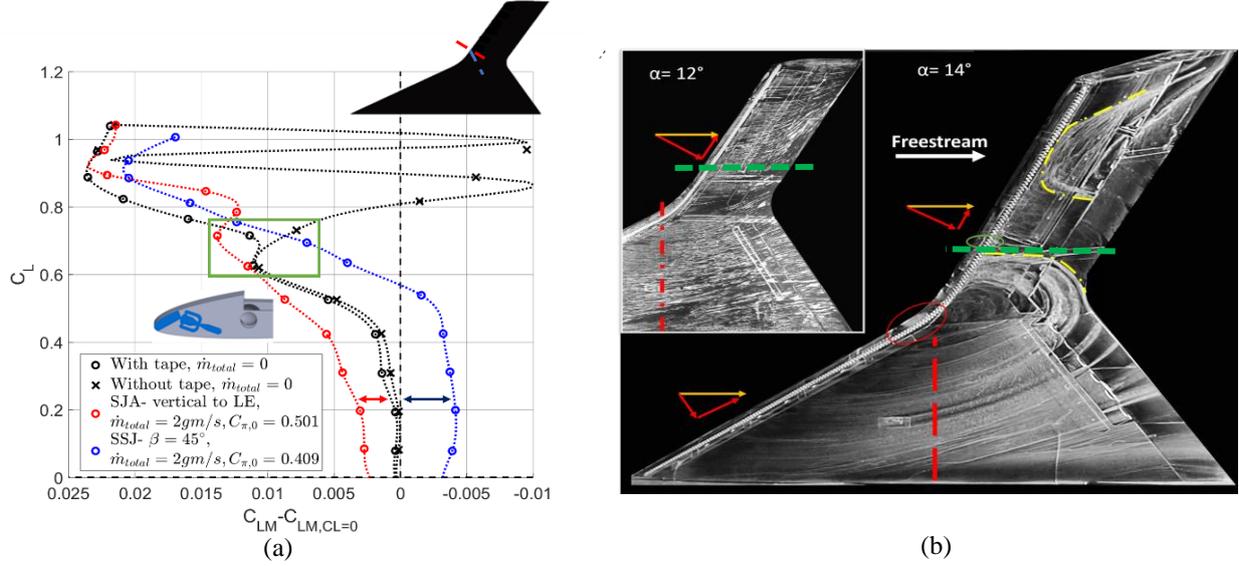

Figure 6: (a) Comparison of the effect on $C_L$ vs $C_{LM}$ due to a LE slot by itself, with a sweeping jet actuator (SJA) at the LE, and a supersonic steady jet (SSJ) oriented at $\beta=45°$ to LE; (b) LEV lift-off and its entrainment and separation effect on the outer wing for a) $\alpha=12°$ (inset) and $\alpha=14°$.

Uncovering a minute rectangular nozzle (1.27mm thick and 2mm wide) near the LE of the outer wing (just outboard of the crank, location is highlighted by a green oval in figure 6-b) significantly impacts the $C_{LM}$ behaviour at $C_L >0.6$ (black curve with 'x' markers compared with black curve with 'o' markers in figure 6-a). It is hypothesized to redirect the LEV downstream preventing its lift-off, generating nose-down pitch moment. This effect could be negated by using sweeping jet oriented vertically from this slot, serving as a fluidic fence. It encourages the LEV to lift-off and generate a nose-up pitch moment. Additionally, a steady supersonic jet actuator (SSJ) employed at the same location but inclined $\beta=45°$ to the outer wing's LE consistently generated a lower $C_{LM}$ value than the baseline case (blue curve with 'o' markers). It also avoided the intermittent nose-down and back up behaviour noticed for the baseline case between $\alpha=12°$-$14°$. It is evident that by manipulating the LEV trajectory, one can achieve an "on-demand" nose-up or nose-down pitch moment.

Stereoscopic Particle Image Velocimetry (PIV) was employed just outboard of the crank ($z/b=0.55$) to examine the ensemble averaged in-plane velocity magnitude ($\sqrt{\overline{U}^2 + \overline{V}^2}/U_\infty$) with streamlines, and out-of-plane velocity ($\overline{W}/U_\infty$) for baseline flow at $\alpha=12°$ (figure 7- a, b), & at $\alpha=14°$ (c, d), and compare it to active flow control (AFC) flow at $\alpha=14°$ (e, f). At $\alpha=12°$ the in-plane flow is attached, with highly accelerated flow between $0<x/c_{streamwise}<0.4$. This low-pressure region results in spanwise flow to be directed inboard from the measurement plane towards the crank, as observed near the LE in figure 7-b. Therefore, there must be a saddle point outboard of the measurement plane near the LE, because there is a slow outboard flow near the surface from $x/c_{streamwise}> 0.7$. This outboard flow is also visible in the inset of figure 6-b.

Increasing the incidence angle to $\alpha=14°$ reduces the accelerated flow near the LE, and the remainder of the wing is characterized by substantial recirculating region (figure 7-c). Pinpointing the origin of the separation streamline is challenging due to laser light reflections contaminating near-surface PIV data, but it is approximately located at $x/c_{streamwise}$ ~0.15. The out-of-plane contours ($\overline{W}/U_\infty$ figure 7-d) highlights three significant structures near the surface. Near LE, strong negative velocity represents the inboard saddle flow, which changes its sign to outboard velocity at 0.175< $x/c_{streamwise}$ < 0.3). The outboard velocity is associated with the formation of a new LEV on the outer wing. Further downstream, positive outboard flow around 0.5< $x/c_{streamwise}$ <1 corresponds to the outboard surface streamlines seen by the surface oil flow that may represent the remnants of the inboard LEV that lifted-off near the crank (figure 6-b). Notably, the contour region encompassed between $\overline{W}/U_\infty$= -0.2 and 0.1 may signify the inboard LEV undergoing the lift-off process.

With actuation from a single SSJ (highlighted by the solid red oval in the inset to figure 7-f), the ensemble averaged recirculating region was effectively eliminated (figure 7-e). The inset shows the oil-flow visualization capturing the core of the vortex lift-off near the crank (highlighted by broken red oval). The entrainment effect of the LEV lift-off seems to be curtailed to a smaller region by the actuator jet. Small recirculating regions upstream and downstream of the jet are due to its interaction with the surface flow surrounding it. When $\overline{W}/U_\infty$ profiles of the actuated flow are compared with the three notable structures in the baseline flow, the discernible differences emerge for $x/c_{streamwise}$ > 0.4 (figure 7-f). The steady jet is represented by a concentration of inboard velocity between 0.4< $x/c_{streamwise}$ <0.5 and surrounding the jet small regions of outboard flow are noted. Notably, the absence of outboard flow closer to the TE suggests that the extent of the inboard LEV liftoff is limited. This absence mitigates potential interactions between the primary and the newly formed outer wing LEV, which originally dominated the recirculating region in the baseline flow. The ensemble averaged results highlight the significant impact of the jet on the flow dynamics and the consequent alteration of key flow structures, but the dynamics of the interaction remains to be investigated.

Using a high-speed camera and a continuous wave laser light sheet aligned approximately parallel to the outer wing at $\alpha$=14°, preliminary smoke-flow visualization revealed flow unsteadiness. Proper Orthogonal Decomposition (POD) technique was implemented on the pixel intensities of the smoke concentration in the images, and the phasic reconstruction of the analysis focusing on the outer wing is presented in figure 8. The smoke impinges the outer wing just downstream of the crank and follows the streamwise direction for $\Theta$=0°. As phase progression occurs, a secondary structure forms, detaching and flowing over the outer wing. Although qualitative, these results indicate the periodicity in the outer wing baseline flow-field, underscoring the importance of investigating the unsteadiness associated with the recirculating region in figure 7-c. Notably for the actuated case, this dynamic behaviour was drastically attenuated, corroborating with attached flow observations in figure 7-e.

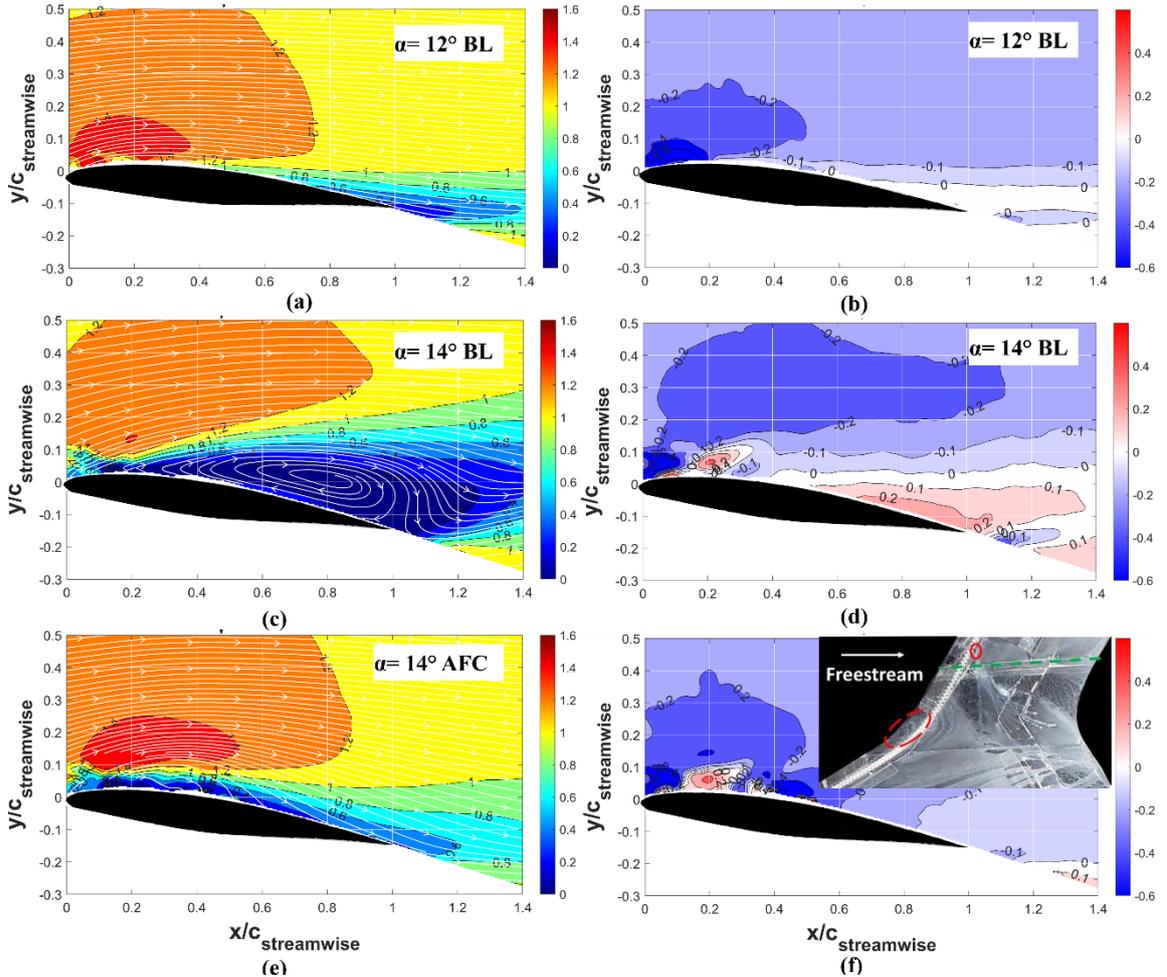

Figure 7: Ensemble averaged (a, c, e) of in-plane velocity magnitude $(\sqrt{\overline{U}^2 + \overline{V}^2}/U_\infty)$ with streamlines, and (b, d, f) out-of-plane velocity $(\overline{W}/U_\infty)$ for (a, b) BL at $\alpha=12°$, (c, d) BL at $\alpha=14°$, and (e, f) AFC at $\alpha=14°$, at $z/b$=0.55.

To quantitatively analyze the unsteady behaviour, POD was performed on a thousand PIV snapshots. Figure 9-a shows the Turbulent Kinetic Energy (% TKE) for the baseline flow for $\alpha$=14° at $z/b$=0.55. Mode 1 ($\Phi_1$) has the highest TKE value of ~14%, and the total energy in the first 10 modes amount to 45%. The contour plots of first 4 spatial modes associated with $U/U_\infty$ are plotted in figure 9-c. Mode 1 marks the unsteadiness in the onset of the recirculating region oscillating between 0.2< $x/c_{streamwise}$ <0.6 (separation streamline). Modes 2-4 highlight the alternating convective structures evolving closer to TE and downstream of the wing. Although not shown here, but $V/U_\infty$ (y-normal velocity component) and $W/U_\infty$ (out-of-plane velocity component) also showed similar convective structures for modes 2-4. Cross-plotting the temporal coefficients of mode 1 and 2 in figure 9-b aids in interpreting the phase correlation between the separation streamline unsteadiness and the development of the convective structures. The plot resembles a circular disk with the mean denoted by a red dot. This highlights the periodicity in the interactions of the underlying modes. Each location is characterized by a unique phase value, enabling reconstruction of different phases for a comprehensive understanding of flow field evolution.

Figure 10 plots the phasic reconstruction of the summation of ensemble average and the first 10 modes, achieved by averaging the temporal coefficients within a wedge of ±5° at four locations across the phase portrait (figure 9-c). Figure 10-a, plots the phasic reconstruction of in-plane velocity magnitude ($\sqrt{U^2 + V^2}/U_\infty$) with streamlines and figure 10-b, plots the out-of-plane velocity ($W/U_\infty$) for the baseline flow at $\alpha$=14°. For the $\Theta$=0° reconstruction, the separation streamline approximately follows the wing curvature until it reconnects with the wing surface near the TE- the separation bubble is shallow. The $W/U_\infty$ reconstruction reveals a concentration of outboard velocity at $x/c_{streamwise}$ ~0.2 representing the new LEV structure on the outer-wing, and positive outboard flow for $x/c_{streamwise}$ >0.7 as observed in the oil-flow viz. (figure 6-b). For the $\Theta$=90° reconstruction, a noticeable expansion of the recirculating region in the in-plane velocity contours is observed. This enlargement is facilitated by a significant increase in $W/U_\infty$ concentration for the new LEV (0.1<$x/c_{streamwise}$<0.4). Additionally, there is a shift in the zero-velocity boundary of the outboard flow, moving upstream from $x/c_{streamwise}$>0.5 to ~0.3, suggesting a potential interaction of the new LEV with the outboard flow originating from the inboard LEV. The largest recirculating region among those investigated is observed for $\Theta=180°$ reconstruction, with the surface streamlines forming a saddle and a nodal point of separation downstream of the wing at $x/c_{streamwise}$ ~1.2. The corresponding $W/U_\infty$ reconstruction highlights high intensity of outboard flow over majority of the wing. At the final phase investigated ($\Theta=270°$), the in-plane velocity magnitudes reveal a separated structure beyond the TE, with its nodal point of separation at $(x, y)/c_{streamwise}$ ~ (1.2, -0.05), downstream of the wing section. The separation streamline is reconnected to the wing surface near TE. Large $W/U_\infty$ outflow is noted beyond $x/c_{streamwise}$ >0.5, but a relatively small concentration exists for the new LEV at $x/c_{streamwise}$ ~0.2, indicating completion of the interaction. This dynamic behaviour of the LEV created on the inboard wing and lifted off by the crank, and its potential interaction with the newly created LEV on the outer wing, constitutes the central focus of the current research.

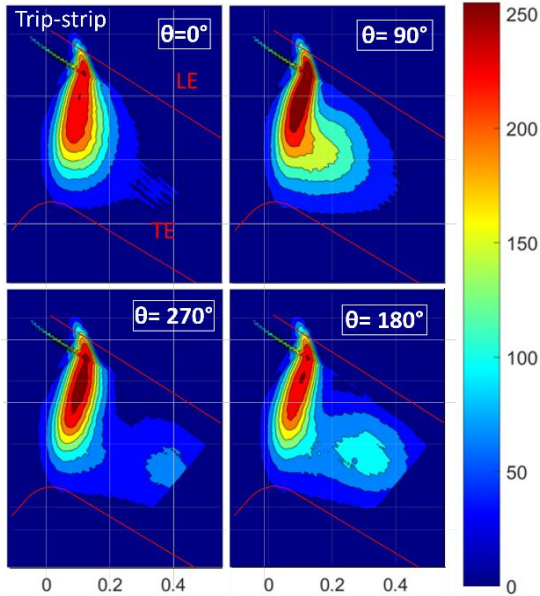

Figure 8: POD analysis of smoke flow visualization images.

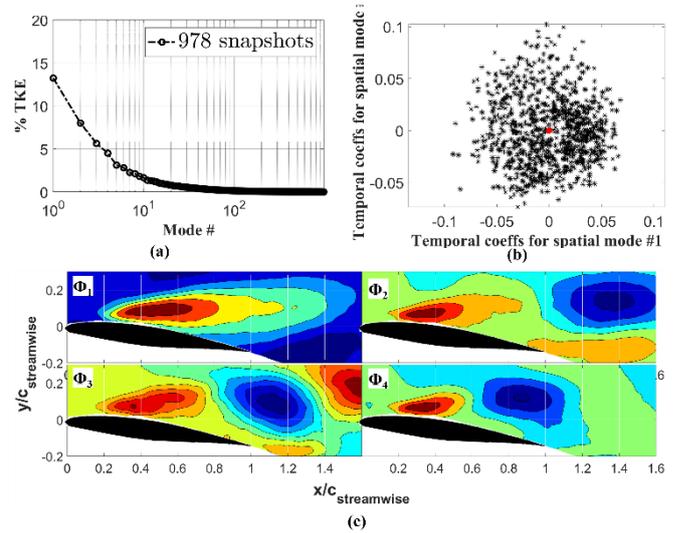

Figure 9: POD analysis of baseline flow at $\alpha$=14°, $z/b$=0.55 detailing a) % TKE, b) phase portrait of spatial modes $\Phi_1$ and $\Phi_2$, and c) contour plots of spatial modes $\Phi_1$, $\Phi_2$, $\Phi_3$ and $\Phi_4$ of $U/U_\infty$.

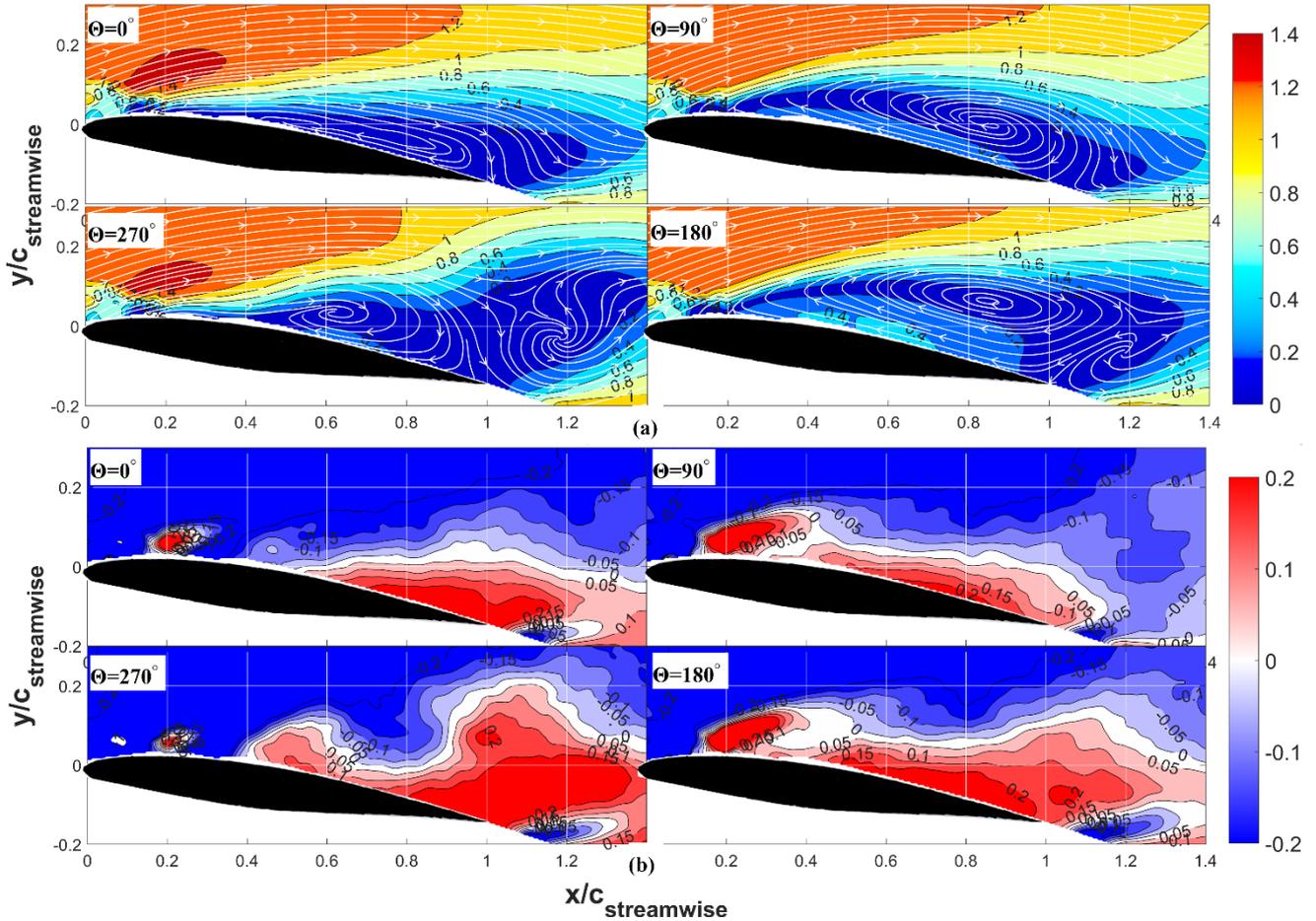

Figure 10: Phasic reconstruction of first 10 spatial modes with ensemble average of (a) in-plane velocity magnitude ($\sqrt{U^2 + V^2}/U_\infty$) with streamlines and (b) out-of-plane velocity ($W/U_\infty$) for baseline flow at $\alpha=14°$ and $z/b=0.55$.

Examining the actuated case in figure 11 reveals the dynamic behaviour but with an attenuated magnitude, primarily confined to the downstream region of the jet location ($x/c_{streamwise}$ >0.5). Traces of recirculating region near the surface can be observed at $\Theta=90°$ and $180°$ for the in-plane velocity contours (figure 11-a), and the associated structures in the $W/U_\infty$ profiles (figure 11-b). The outer wing LEV maintains its center at $x/c_{streamwise}$ ~0.2, displaying a relatively consistent-sized structure across all the phases. This consistency suggests a lack of interaction with the downstream structures. Downstream of the jet and its recirculating regions, a small pocket of near-zero outboard flow is observed at $0.1<x/c_{streamwise}<0.4$ for $\Theta=0°$, evolving into a larger structure at TE for the $\Theta=180°$ case. This implies that, even in the actuated case, some periodicity is associated with the inboard LEV's entrained flow, albeit at a substantially reduced magnitude.

## 3. INTERIM CONCLUSION

Flow visualization by smoke that was illuminated by a plane laser sheet parallel to the wing's surface at $\alpha=14°$ and using a high-speed photography, revealed flow unsteadiness on the outer wing. Its origin might come from the unsteady K-H type eddies in the interior of the inboard LEV. The unsteadiness persists even after the lift-off of this LEV, and it seems to have affected the newly created LEV outboard of the crank. Proper orthogonal decomposition helped to map the evolution of these unsteady structures that were also seen using PIV. A steady supersonic jet (SSJ) emanating from a nozzle located on the outboard wing and inclined at an inboard direction of $\beta=45°$ to the LE mitigated the separation in the crank region and greatly affected the pitch of the model.

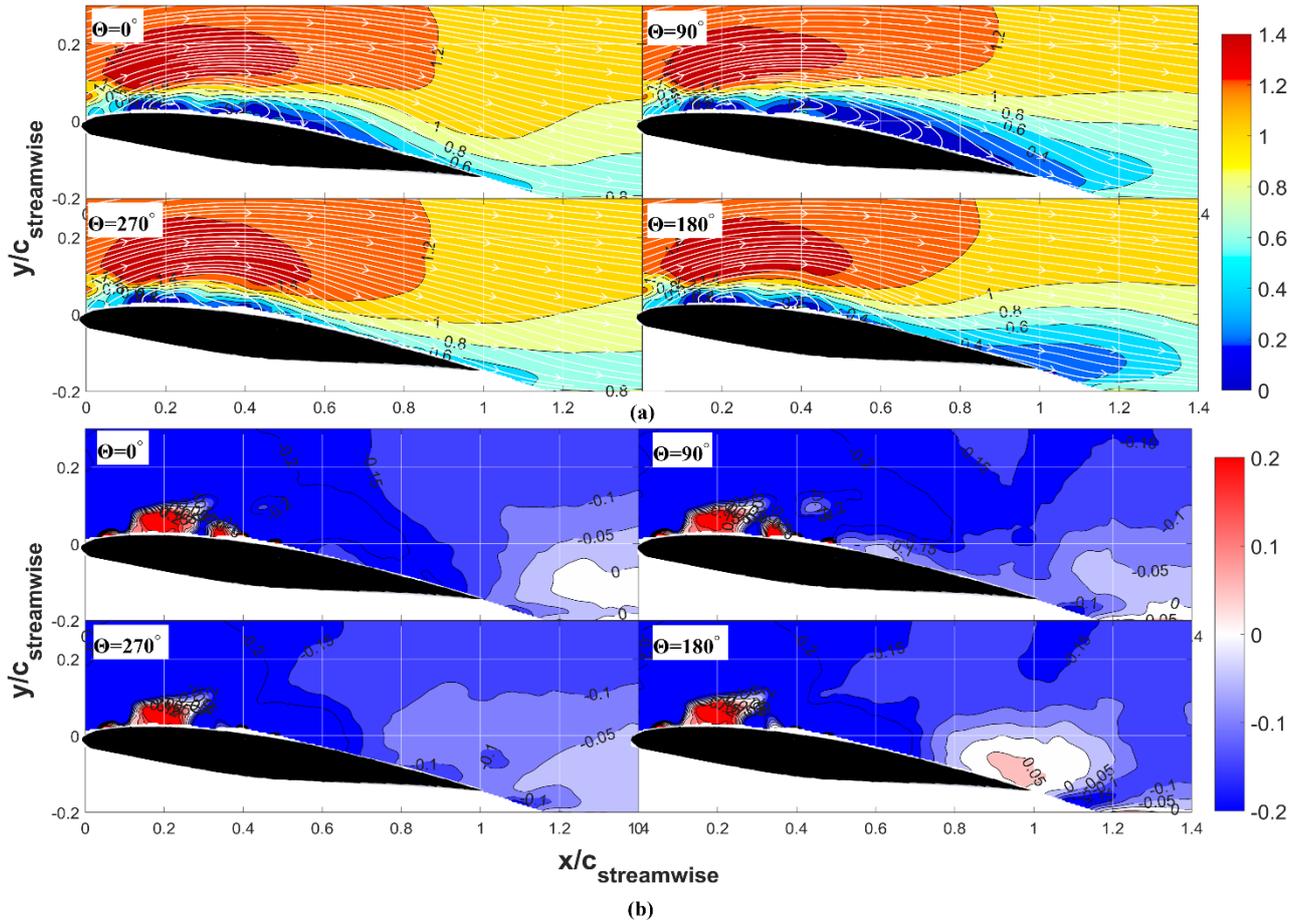

Figure 11: Phasic reconstruction of first 10 spatial modes with ensemble average of (top) in-plane velocity magnitude ($\sqrt{U^2+V^2}/U_\infty$) with streamlines and (bottom) out-of-plane velocity ($W/U_\infty$) for AFC case at $\alpha=14°$ and $z/b=0.55$.

The current work in progress raises more questions than it answers. For example: Is there a single or multiple frequencies dominating the inboard LEV development; and how do these frequencies affect the lift-off process, and ultimately impact the development of the newly created outboard LEV. Finally, how AFC affects the lifted LEV either by changing its structure or its path.